\documentstyle[12pt]{article}
\begin{document}
\vspace*{-1in}
\renewcommand{\thefootnote}{\fnsymbol{footnote}}
\begin{flushright}
CERN-TH/95-118\\
CUPP-95/2\\
hep-ph/9505356 \\
\end{flushright}
\vskip 53pt
\begin{center}
{\Large{\bf \boldmath Constraining the charged Higgs mass
 in the left--right symmetric model from $b \rightarrow s \gamma$}}
\vskip 30pt
{\bf Gautam Bhattacharyya\footnote{gautam@cernvm.cern.ch}
\vskip 10pt
 Theory Division, CERN, \\ CH-1211  Geneva 23,  Switzerland}
\vskip 10pt
and
\vskip 10pt
{\bf Amitava Raychaudhuri\footnote{amitava@cubmb.ernet.in}
\vskip 10pt
Department of Pure Physics, University of Calcutta, \\
92 Acharya Prafulla Chandra Road, Calcutta 700 009, India.}
\vskip 70pt

{\bf ABSTRACT}
\end{center}

In the context of the left--right symmetric model,
the decay $b \rightarrow
s \gamma$ receives contributions from
the gauge interactions mediated mainly by the $W_L$, through
$W_L$--$W_R$ mixing and also from
the Yukawa interactions of
the charged and the neutral (flavour-changing) scalars
(the latter type of Yukawa interaction has been overlooked
in the previous literature).
Following the recent CLEO measurement of the inclusive
$b \rightarrow s \gamma$ process
and the measurement of the top-quark mass
by the CDF and D0 collaborations, the parameter space of the
left--right symmetric model is more squeezed than before.

\vskip 53pt
\begin{center}
             {\it Phys. Lett.} {\bf B357} (1995) 119.
\end{center}
\begin{flushleft}
CERN-TH/95-118\\
May 1995\\
\end{flushleft}

\setcounter{footnote}{0}
\renewcommand{\thefootnote}{\arabic{footnote}}
\vfill
\clearpage
\setcounter{page}{1}
\pagestyle{plain}

In the $SU(2)_L \otimes SU(2)_R \otimes U(1)$
left--right symmetric model (LRM)  \cite{lrm} there is a
new scale at which the gauge group breaks to the
$SU(2)_L \otimes U(1)$ Standard Model (SM).
The sensitivity to this scale of low-energy phenomena such as
$K$--$\bar{K}$ mixing and neutrino masses \cite{led} has been a
subject of wide interest over the past few years.
Of late, a particularly interesting channel to examine
various species of new physics, including this
LRM scenario, has been provided
by the inclusive $B$-decay measurement by the CLEO collaboration,
$B(b\rightarrow s \gamma) = (2.32 \pm 0.57 \pm 0.35)
\times 10^{-4} \rightarrow (1.0-4.2) \times 10^{-4}$ (at 95\% C.L.)
\cite{cleo}. It has already
been pointed out (\cite{coc}--\cite{rizzo})
that this rare decay has a strong influence
on restricting the parameter space of the LRM. Recently,
Cho and Misiak \cite{cho} investigated the effects of
$W_L$--$W_R$ mixing on $b \rightarrow s \gamma$ with an extensive
analysis of QCD corrections which are very important for this
process; however, they have not considered the contributions from
the scalar sector. Babu et al. \cite{babu} included
the charged scalars in the analysis,
but their treatment of QCD corrections is incomplete.
In this paper we attempt to improve upon the previous analyses by
\begin{itemize}
\item
 including all the above contributions coherently
in a single analysis,
\item
 incorporating the contribution of the flavour-violating neutral
scalars, which has so far been {\it overlooked}, and
\item
 reexamining the parameter space in the light of the  new  CLEO
measurement of the $b \rightarrow s \gamma$ inclusive branching
ratio \cite{cleo} and the recent CDF and D0 measurements
of the top-quark mass as $m_t = 180 \pm 12$ GeV
\cite{top} \footnote{The value is the weighted average of
$m_t = 176 \pm 13$ GeV (CDF) and $m_t = 199 \pm 30$ GeV (D0).}.
\end{itemize}

\vskip 9pt
In the $SU(2)_L \otimes SU(2)_R \otimes U(1)$ gauge model
the quarks ($q$) and the leptons ($l$)
transform as $q_L (2,1,1/3)$,~$q_R (1,2,1/3)$,~
$l_L (2,1,-1)$,~and $l_R (1,2,-1)$.~The scalar sector consists of the
following Higgs fields:~ $\Delta_L (3,1,2)$, $\Delta_R (1,3,2)$ and
$\Phi (2,2,0)$, of which only the latter participates in the Yukawa
interaction and is explicitly shown as:
\begin{equation}
\Phi = \pmatrix{~\phi_1^0 & \phi_2^+ \cr~\phi_1^- & \phi_2^0}.
\end{equation}
$\Delta_R$ is used to break $SU(2)_L\otimes SU(2)_R
\otimes U(1)_{B-L}$
to $SU(2)_L \otimes U(1)$, while $\Delta_L$ is introduced to
maintain a discrete parity invariance. The vev  $v_R$
of $\Delta_R$ sets
the LRM breaking scale. The vevs of $\Phi$ are given by
$\langle\phi_1^0\rangle = k$ and $\langle\phi_2^0\rangle = k'$.
One requires $v_R \gg k,k'$ from
the excellent agreement between the $(V-A)$ theory
and the experimental data, while the
hierarchy $v_L \ll k,k'$ is set from the $\rho$ parameter constraint.
The SM gauge group is reproduced in the limit $v_R \rightarrow
\infty, v_L \rightarrow 0$. We neglect any small phase difference
between $k$ and $k'$.

\vskip 9pt
Introducing $\tan\beta = k/k'$, we obtain the physical charged
scalars and the second set of neutral scalar and pseudoscalar
\footnote{The neutral scalar and the pseudoscalar in the first set
are identified with the  Higgs and the longitudinal component
of the lighter $Z$, respectively, whose Yukawa couplings are
flavour-diagonal.}
as
\begin {eqnarray}
H^\pm & = & \cos\beta~ \phi_1^\pm + \sin\beta~ \phi_2^\pm,
\nonumber \\
H_2^0 & = & \sqrt{2} \left(- \sin\beta~ {\rm Re} \phi_2^0 + \cos\beta
{}~{\rm Re} \phi_1^0 \right), \\
G_2^0 & = & \sqrt{2} \left( \cos\beta~ {\rm Im} \phi_1^0 + \sin\beta
{}~{\rm Im} \phi_2^0 \right) \nonumber.
\end{eqnarray}
The Yukawa interaction in the quark sector is given by the Lagrangian
\cite{babu} \footnote{We use the same notation as in \cite{babu}
as much as possible.}:
\begin{equation}
L_Y = \bar{q}_L h \Phi q_R + \bar{q}_L \tilde{h} \tilde{\Phi} q_R
+ {\rm h.c.}
\end{equation}
where $\tilde{\Phi} \equiv \tau_2 \Phi^* \tau_2$ and $h, \tilde{h}$
are $3 \times 3$ Hermitian matrices in flavour space.
The up- and down-type quark mass matrices are given by:
\begin{eqnarray}
M_u & = & h k + \tilde{h} k',\nonumber \\
M_d & = & h k' + \tilde{h} k.
\end{eqnarray}

\vskip 9pt

After a straightforward calculation, the charged and neutral current
(which are relevant to $b \rightarrow s \gamma$) Yukawa Lagrangian,
in a more transparent form, are given by:
\begin{eqnarray}
L_Y^C & = & -{N\over {\cos{2\beta}}} \bar{u}_i \left[\sin{2\beta}
\left(\hat{M}_u V P_L - V \hat{M}_d P_R\right) \right. \nonumber \\
 &+ & \left.\left(\hat{M}_u V P_R - V \hat{M}_d P_L\right)  \right]
H^+  d_i + {\rm h.c.}
\end{eqnarray}
and
\begin{equation}
L_Y^N (d)= {N\over \sqrt{2}\cos{2\beta}} \bar{d}_i \left[
(V^\dagger\hat{M}_u V)_{ij} - \sin{2\beta} \hat{M}_d
\delta_{ij}\right] \left(H_2^0 - i \gamma_5 G_2^0\right) d_j,
\end{equation}
where $\hat{M}_u$ and $\hat{M}_d$ are diagonal up- and down-type
quark mass matrices, $V$ is the standard Cabibbo-Kobayashi-Maskawa
mixing matrix and $N = 1/\sqrt{k^2 + k'^2}$.

\vskip 9pt

The contribution of the flavour-violating neutral scalars
has been overlooked in the previous
literature, which we find to be significant for
some region of the parameter space.
It may be noted that $H_2^0$ and $G_2^0$ mediate flavour violation
even in the limit $\beta \rightarrow 0$. This originates from the
fact that the Yukawa interaction
mediated by the bidoublet scalar in LRM does not reproduce the
SM scenario in the above limit.

\vskip 9pt
In the charged gauge boson sector, $W_L^\pm$ and $W_R^\pm$ mix
to give the lighter and heavier mass eigenstates as \footnote{Indeed,
$W_1^\pm$are identified with the $W^\pm$ of the SM.}
\begin{eqnarray}
W_1^\pm & = & \cos\xi~W_L^\pm + \sin\xi~W_R^\pm, \nonumber \\
W_2^\pm & = & -\sin\xi~W_L^\pm + \cos\xi~W_R^\pm.
\end{eqnarray}
where the mixing angle $\xi$ is given by
\begin{equation}
\xi \simeq \sin{2\beta} {{m^2_{W_1}}\over {m^2_{W_2}}}
\equiv \sin{2\beta} {{m^2_W}\over {m^2_{W_2}}}.
\end{equation}

\vskip 9pt
The effective Lagrangian relevant for the process
$b\rightarrow s \gamma$ can be
written as \footnote{We neglect the contributions proportional
to $m_s$.}:
\begin{eqnarray}
{\cal L}_{eff} & = &{\sqrt{{G^2_F}\over{8\pi^3}}}
 V_{tb} V^*_{ts}~m_b
\left[\sqrt{\alpha}\left\{A_{\gamma L}~
\overline{s}_L\sigma^{\mu\nu} b_R
+ A_{\gamma R}~\overline{s}_R \sigma^{\mu\nu} b_L\right\}
 F_{\mu\nu}\right.
\nonumber \\
& + & \left. \sqrt{\alpha_S} \left\{A_{gL}~ \overline{s}_L
T_a \sigma^{\mu\nu} b_R
+ A_{gR}~ \overline{s}_R T_a \sigma^{\mu\nu} b_L\right\}
  G^a_{\mu\nu}\right] + {\rm h.c.},
\label{heff}
\end{eqnarray}
\noindent
where the contributions to $A_{\gamma L}, A_{\gamma R}, A_{gL}$
and $A_{gR}$ from the different sectors
\footnote{We do not show, in the expressions for $A_{\gamma R}$
and $A_{gR}$, the $W_2$-induced contributions which are sufficiently
damped since $m_{W_2}$ is set to 1.6 TeV \cite{languma} all along our
analysis. However, we have included it in our numerical code.}
are given by ($x = m_t^2/m_W^2 , y =
m_t^2/m^2_{H^+}, z_1 = m_b^2/m^2_{H_2^0}, z_2 = m_b^2/m^2_{G_2^0}$)
\footnote{Whenever we encounter a $b$-quark inside a triangle
 loop, we take
the running mass $m_b (m_W) \sim 3$ GeV, as in \cite{babu}.}:
\begin{eqnarray}
A_{\gamma L} & = & A_{\gamma}^{\rm SM} (x) + \xi {m_t\over {m_b}}
A_\gamma^{\rm mix} (x) + \tan^2{2\beta}~ F_\gamma(y) +
{{m_t \sin{2 \beta}}
\over {m_b \cos^2{2\beta}}}~ G_\gamma(y) \nonumber \\
& + & {{Q_b}\over 4} {m_t \over
{m^2_{H_2^0}}} \left({m_t\over {\cos^2{2\beta}}} -
{{m_b \sin{2\beta}} \over{\cos^2{2\beta}}}\right)
\left\{H(z_1) + G(z_2)\right\}, \nonumber \\
A_{\gamma R} & = & \xi {m_t\over {m_b}}
A_\gamma^{\rm mix} (x) + \sec^2{2\beta}~
F_\gamma(y) + {{m_t \sin{2 \beta}}
\over {m_b \cos^2{2\beta}}}~ G_\gamma(y) \nonumber \\
& + & {{Q_b}\over 4} {m_t \over
{m^2_{H_2^0}}} \left({m_t\over {\cos^2{2\beta}}} -
{{m_b \sin{2\beta}} \over {\cos^2{2\beta}}}\right)
\left\{H(z_1) + G(z_2)\right\} ,\\
A_{gL} & = & A_{g}^{\rm SM} (x) + \xi {m_t\over {m_b}}
A_g^{\rm mix} (x) + \tan^2{2\beta}~ F_g(y) + {{m_t \sin{2 \beta}}
\over {m_b \cos^2{2\beta}}}~ G_g(y) \nonumber \\
& + & {1\over 4} {m_t \over
{m^2_{H_2^0}}} \left({m_t\over {\cos^2{2\beta}}} -
{{m_b \sin{2\beta}} \over {\cos^2{2\beta}}}\right)
\left\{H(z_1) + G(z_2)\right\} \nonumber, \\
A_{gR} & = & \xi {m_t\over {m_b}}
A_g^{\rm mix} (x) + \sec^2{2\beta}~ F_g(y) + {{m_t \sin{2 \beta}}
\over {m_b \cos^2{2\beta}}}~ G_g(y) \nonumber \\
& + & {1\over 4} {m_t \over
{m^2_{H_2^0}}} \left({m_t\over {\cos^2{2\beta}}} -
{{m_b \sin{2\beta}} \over {\cos^2{2\beta}}}\right)
\left\{H(z_1) + G(z_2)\right\}. \nonumber
\end{eqnarray}
The functions in the above expressions are given by
\footnote{$A_\gamma^{\rm SM}$ and $A_g^{\rm SM}$ were first
calculated by Inami and Lim \cite{inamilim}.}
\begin{eqnarray}
A_\gamma^{\rm SM} & = & {{x(8x^2+5x-7)}\over{24(1-x)^3}}
+ {{x^2(3x-2)}\over{4(1-x)^4}}\ln x, \nonumber \\
A_g^{\rm SM} & = & {{x(x^2-5x-2)}\over{8(1-x)^3}}
-{{3x^2}\over{4(1-x)^4}}\ln x, \nonumber \\
A_\gamma^{\rm mix} & = & {{-20+31x-5x^2}\over{12(1-x)^2}}
+ {{x(3x-2)}\over{2(1-x)^3}}\ln x, \nonumber \\
A_g^{\rm mix} & = & -{{4+x+x^2}\over{4(1-x)^2}} - {{3x}
\over{2(1-x)^3}} \ln x, \nonumber \\
F_\gamma (x) & = & {x(-25+53x-22x^2) \over {72(1-x)^3}}  +
{x^2(2-x)\over {12(1-x)^4}} \ln x, \\
G_\gamma (x) & = & {x(5x-3) \over {12(1-x)^2}}  - {{x(2-3x)}
\over {6(1-x)^3}} \ln x, \nonumber \\
F_g (x) & = & {-20x+19x^2-5x^3\over {24(1-x)^3}} -
{x\over {4(1-x)^4}} \ln x, \nonumber \\
G_g (x) & = & {-3x+x^2 \over {4(1-x)^2}} -
{x\over {2(1-x)^3}} \ln x, \nonumber \\
H(x) & = & {{16-29x+7x^2}\over{12(1-x)^3}}
+ {{2-3x}\over{2(1-x)^4}}\ln x ,\nonumber \\
G(x) & = & {{20-19x+5x^2}\over{12(1-x)^3}}
+ {{2-x}\over{2(1-x)^4}}\ln x. \nonumber
\end{eqnarray}

\vskip 9pt

It should be noted that there is a {\it disagreement of sign} between
Babu et al.'s \cite{babu} and Asatryan et al.'s \cite{asatryan}
calculation of the charged scalar-induced contribution.
Our calculation {\it agrees with the latter}.

\vskip 9pt

The effective Wilson coefficients can now be written as:
\begin{eqnarray}
\label{cleff}
C_{7L}^{\rm eff} &=
&\eta^{16/23} A_{\gamma L} + {8\over 3} (\eta^{14/23} -
\eta^{16/23}) A_{gL} + C + C',  \\
\label{creff}
C_{7R}^{\rm eff} &=
&\eta^{16/23} A_{\gamma R} + {8\over 3} (\eta^{14/23} -
\eta^{16/23}) A_{gR} + C'.
\end{eqnarray}
In the above equations, $\eta = \alpha_S(M_Z)/\alpha_S(\mu)$,
where $\mu$ is the QCD renormalization scale; $C$ corresponds to
the leading log QCD corrections in SM \cite{ciu}, and $C'$
refers to the extra contribution from mixing of
additional operators in LRM, which has been computed in
ref. \cite{cho}.
These are given by
\begin{equation}
C = \sum_{i=1}^8 h_i \eta^{a_i},
\end{equation}
where
\begin{equation}
a_i = \frac{14}{23}, \frac{16}{23}, \frac{6}{23}, -\frac{12}{23},
      0.4086, -0.4230, -0.8994, 0.1456
\end{equation}
\begin{equation}
h_i = \frac{626126}{272277}, -\frac{56281}{51730},
-\frac{3}{7}, -\frac{1}{14}, -0.6494, -0.0380, -0.0186, -0.0057
\end{equation}
and
\begin{equation}
C' = \xi {{m_c}\over {m_b}} \sum_{i=1}^4 h'_i \eta^{a'_i},
\end{equation}
where
\begin{equation}
a'_i = (0.6957, 0.6087, -1.0435, 0.1304)
\end{equation}
\begin{equation}
h'_i = (-0.6615, 1.3142, 0.0070, 1.0070).
\end{equation}

\vskip 9pt

Finally, the branching ratio of $b\rightarrow s\gamma$ is given,
in units of the semileptonic $b$-decay branching ratio, by
\begin{equation}
 {{B (b\rightarrow s\gamma)} \over{B
(b\rightarrow ce\overline{\nu})}} = {{6 \alpha}\over{\pi\rho\lambda}}
\left|{{V_{tb} V_{ts}^*}\over{V_{bc}}}\right|^2
\left[ |C_{7L}^{\rm eff}|^2 + |C_{7R}^{\rm eff}|^2\right],
\label{brat}
\end{equation}
where
$\rho = (1-8r^2+8r^6-r^8-24r^4 \mbox{\rm ln}r)$
 with $r = m_c/m_b$ and $\lambda =
1 - 1.61\;\alpha_S(m_b)/\pi$.
It may be noted that the ${m_b}^5$ dependence in
the partial decay widths of the $b$ quark cancels out in
eq. (\ref{brat}). An ${\cal{O}}(m^2_s/m^2_b)$ part in the
branching ratio is neglected.
We take $B(b\rightarrow ce\overline{\nu}) = 0.107$.

\vskip 9pt

In Figs. 1--3  we have plotted the branching ratio $B (b \rightarrow
s \gamma)$ as a function of $m_{H^+}$.
We have fixed $m_{W_2} = 1.6$ TeV.
To demonstrate the magnitude of the $H_2^0$-induced effect, we
work with two representative values: $m_{H_2^0} = 300$
GeV and 800 GeV and have displayed their effects
in Figs. 1 and 2, respectively, for $m_t = 180$ GeV.
On the other hand, the chirality-flip of the top
quark inside the loop due to the presence of the right-handed current
is responsible for a strong $m_t$-dependence of our prediction.
To illustrate this point, we demonstrate in Fig. 3 how a line curve
in Fig. 1 corresponding to $\beta = -10^\circ$, as an example,
becomes a thick band due
to the variation of $m_t$ in the range 168--192 GeV.
In Fig. 4 we fix the charged Higgs mass to 800 GeV and plot
the branching ratio $B(b \rightarrow s \gamma)$ as a function
of $m_{H_2^0}$.
The salient features of the relative contributions
of the different sectors of the LRM that emerge from the above
figures are listed below:
\begin{enumerate}
\item
The contributions induced by the $W_L$--$W_R$ mixing and the charged
scalar are very sensitive to the choice of $\beta$.
The reason is that the chirality-flipped $(m_t/m_b)$-enhancement
factor, which constitutes
the potentially largest contribution, multiplies $\sin 2\beta$.
Even a choice of $|\beta| \sim 5^\circ$ can rule out a charged
Higgs up to a mass of several hundred GeV. Evidently, choosing a
larger $|\beta|$ pushes it up even further.

\item
Contrary to the Yukawa couplings of $H_1^0$ or $G_1^0$, those of
$H_2^0$ and $G_2^0$ are {\it not}
totally flavour-diagonal and hence they
contribute to $b \rightarrow s \gamma$. For the sake of simplicity
we have assumed $H_2^0$ and $G_2^0$ to be mass-degenerate.
With increasing $m_{H_2^0}$ the contribution from the neutral
scalar sector decouples fast, which is the origin of the relative
shifts between the curves in Fig. 2 ($m_{H_2^0} = 800$ GeV) and
in Fig. 1 ($m_{H_2^0} = 300$ GeV).
It may be noted, though, that the
individual contributions from  $H_2^0$ and $G_2^0$ are in the same
direction and roughly of the same order of magnitude. So if one
relaxes the condition of their mass degeneracy and their joint
contribution is thought to be of the same order of magnitude as
their individual contributions, the curves should lie somewhere
between their corresponding positions in Fig. 1 and Fig. 2.

\item
In the limit of $\beta = 0^\circ$, the dominant contribution in the
LRM comes from the new operator-mixing effect, represented by $C'$,
and also from the flavour-violating neutral scalar sector
(when those neutral scalar masses are not too heavy).

\item
The $m_t$-dependent (dominant) contributions multiply
$1/(\cos^2 2\beta)$ in the neutral scalar sector and
$\sin 2\beta/(\cos^2 2\beta)$ in the charged scalar sector.
A close look at the relative signs of the associated factors
(see eqs. (10) and (11)) in those contributions reveals that
for a positive (negaive) $\beta$, there is a constructive
(destructive) interference between the effects induced by the
neutral and the charged scalars. The $W_L$--$W_R$ mixing
contribution also depends on the sign of $\beta$ through $\xi$.
A comparison between the curves for $\beta = -5^\circ$ and
$5^\circ$ in Fig. 4, as an example, amply demonstrates this
analytic interplay involving the sign of $\beta$.

\end{enumerate}

\vskip 9pt

At this point, a few words about the theoretical uncertainties
in this process are in order \cite{aligre}.
The evolution of ${{\cal L}_{\rm eff}}$
from $m_W$ to a lower momentum
scale ($\mu \sim m_b$) by the QCD renormalization group analysis
involves a significant theoretical uncertainty regarding a precise
choice of $\mu$ at which $\alpha_S$ is to be determined.
The SM branching ratio for a leading log calculation is quoted as
$B^{\rm SM} (b\rightarrow s \gamma) = (2.8 \pm 0.8) \times 10^{-4}$
\cite{buras} where the error comes mainly from the uncertainty
of $\mu$ in the range $m_b/2 < \mu < 2m_b$.
We take $\mu = m_b$ to obtain $B^{\rm SM} (b\rightarrow
s \gamma) = 2.9 \times 10^{-4}$ for $m_t = 180$ GeV.
Recently a part of the next-to-leading order QCD corrections
has been estimated  with a consequence of reducing the QCD
enhancement in the SM yielding
$B^{\rm SM} (b\rightarrow s \gamma) = (1.9 \pm 0.5) \times 10^{-4}$
\cite{ciu2}.
It may be noted that all our estimates of the LRM contributions
stand above the base value of the SM approximated at the leading
order QCD corrections. If, for instance, the full calculation of the
next-to-leading order QCD corrections pulls the SM estimate down,
the total effect including the LRM contributions will also come down
by the same absolute amount. However, for a non-negligible $\beta$,
the total LRM contribution or even the contribution from each
individual sector as well is shown to be numerically significant
and hence their impact are not likely to be masked by the
uncertainties from the next-to-leading order QCD corrections.

\vskip 9pt

In conclusion: we have investigated the parameter space of the LRM in
the context of $b \rightarrow s \gamma$. Different sectors of the
LRM contributing to this process have been added coherently.
We have included the effect of the flavour-violating neutral scalars,
which was missing in the previous analyses. Upon imposition of the
CLEO measurement of the inclusive $b \rightarrow s\gamma$ rate,
the lower limit of the charged Higgs mass is pushed up to
several hundred GeV even for a small value of $|\beta|$.
Although there are lots of parameters in the LRM which can conspire,
leading to cancellations and reducing the
power of prediction in this model, still a reduction of errors in the
inclusive $b \rightarrow s \gamma$ measurement at CLEO,
a more precise determination of the top-quark mass and, indeed,
a better understanding
of the next-to-leading order QCD corrections will all serve
to constrain the parameter space even more strongly.

\vskip 9pt

AR acknowledges partial support from the Council of Scientific and
Industrial Research and the Department of Science and
Technology, India.

\newpage

\setcounter{figure}{0}
\begin{figure}[htbp]
\vskip 9.0in\relax\noindent\hskip -1in\relax
{\includegraphics{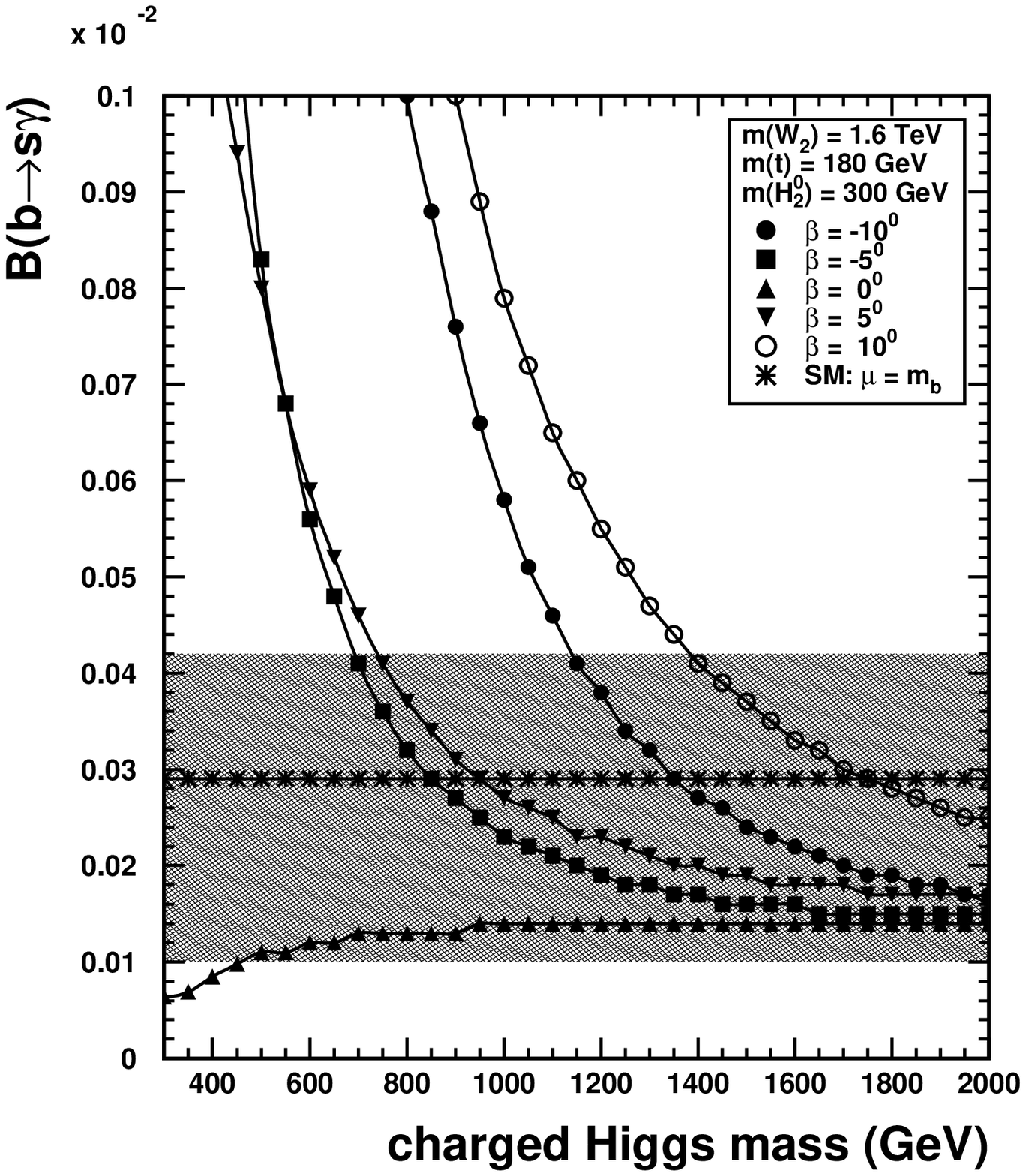}}
\vskip -2in\caption{The branching ratio for the process
$b\rightarrow s\gamma$ as a function of $m_{H^+}$
for different values of $\beta$ in the LRM.
The value of $m_t$ has been fixed to 180 GeV
and $m_{H_2^0} = m_{G_2^0}$ has been set to 300 GeV. The SM line
relies on leading log QCD calculation at $\mu = m_b$. The shaded
area is the experimentally allowed region at 95\% C.L. [3].}
\end{figure}

\newpage

\begin{figure}[htbp]
\vskip 9.0in\relax\noindent\hskip -1in\relax
{\includegraphics{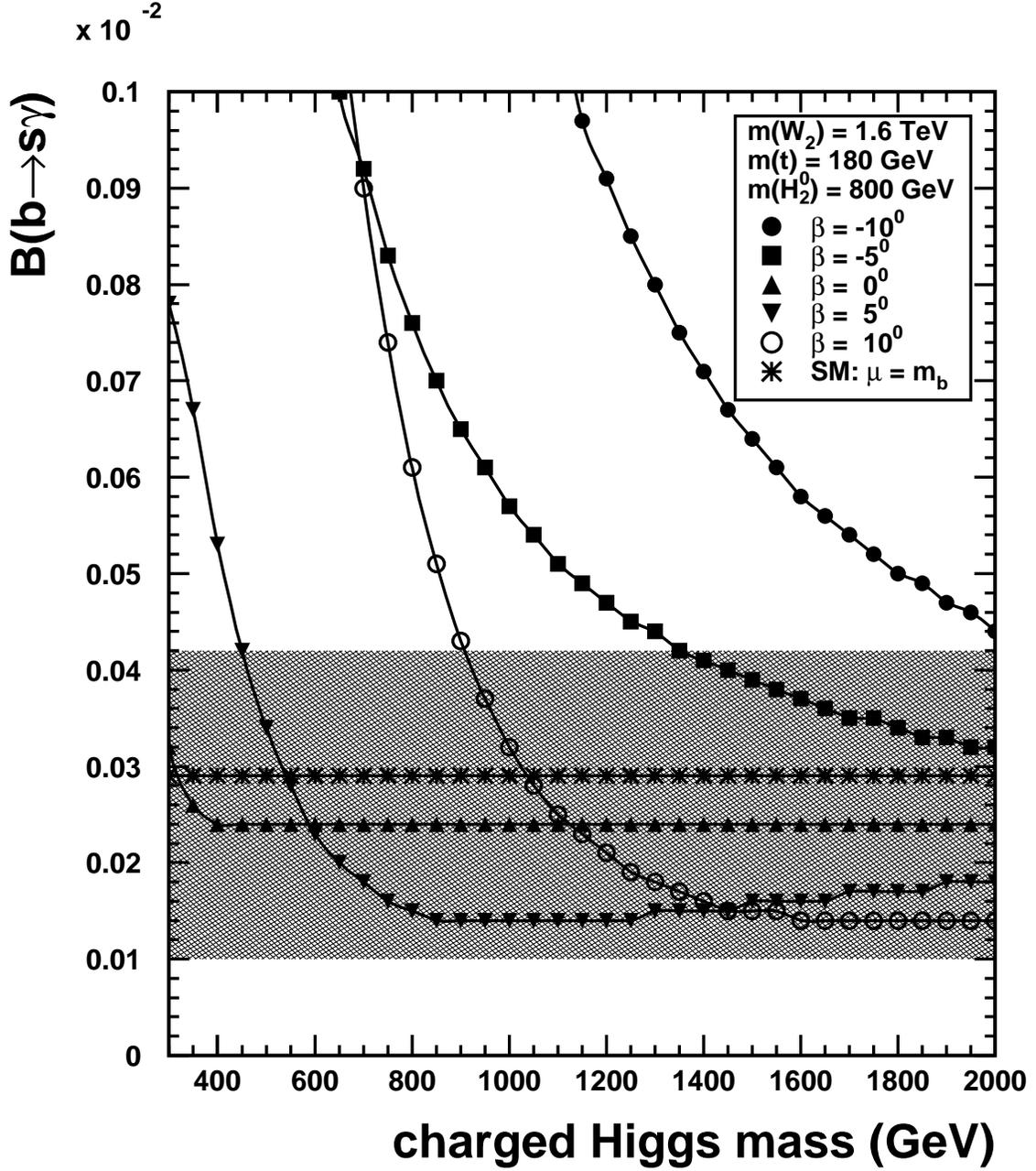}}
\vskip -2in\caption{ Same as in Fig. 1,
but with $m_{H_2^0} = m_{G_2^0} = 800$ GeV.}
\end{figure}

\newpage

\begin{figure}[htbp]
\vskip 9.0in\relax\noindent\hskip -1in\relax
{\includegraphics{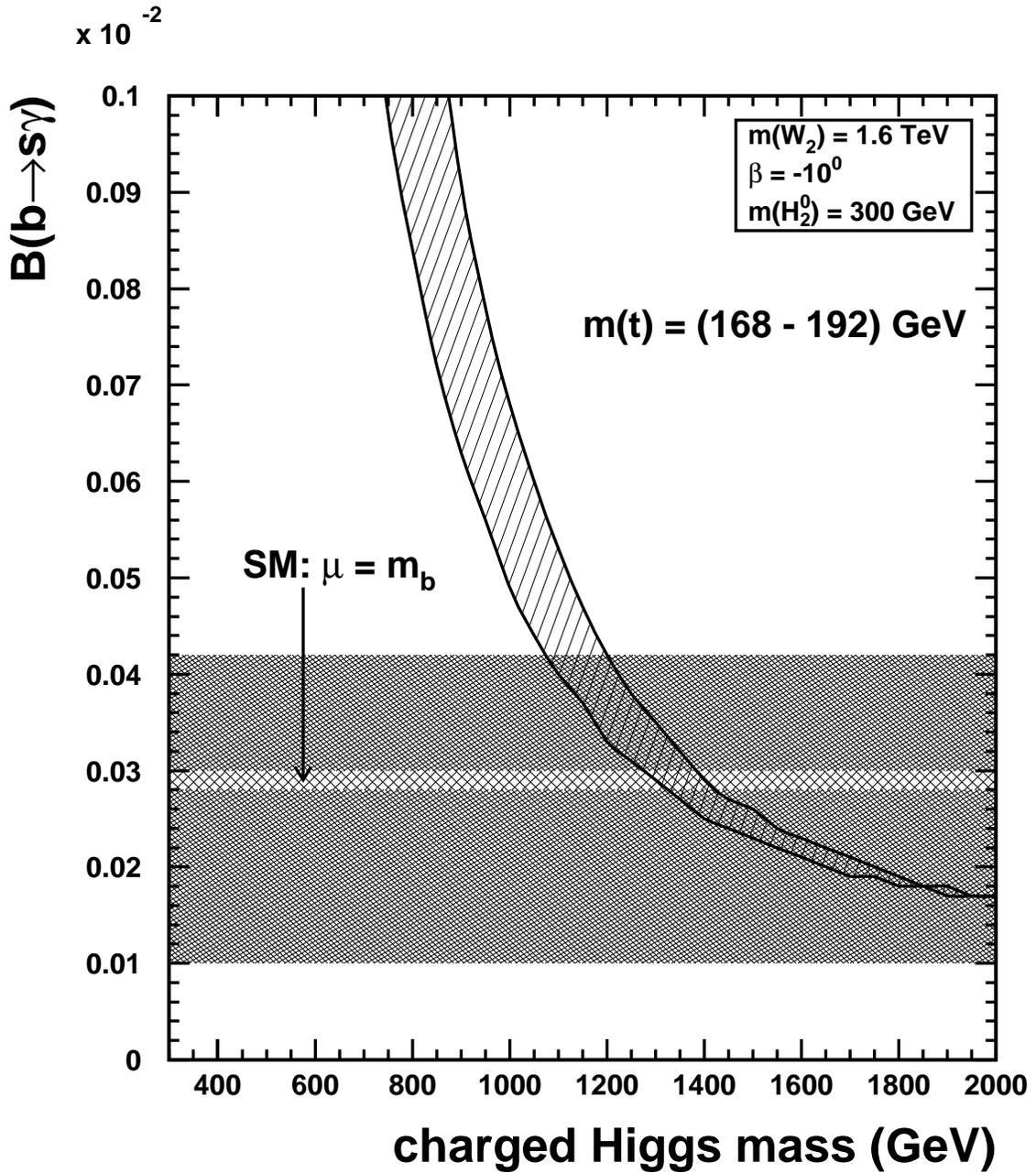}}
\vskip -2in\caption{ The $m_t$ dependence of the result is exhibited
by showing how the curve corresponding to $\beta = -10^\circ$ (as an
example) in Fig. 1 becomes a band (hatched). The SM line for $\mu =
m_b$ also becomes a band (horizontal hatched strip within the shaded
area) due to the same effect. The shaded area is the experimentally
allowed region at 95\% C.L. [3].}
\end{figure}

\newpage

\begin{figure}[htbp]
\vskip 9.0in\relax\noindent\hskip -1in\relax
{\includegraphics{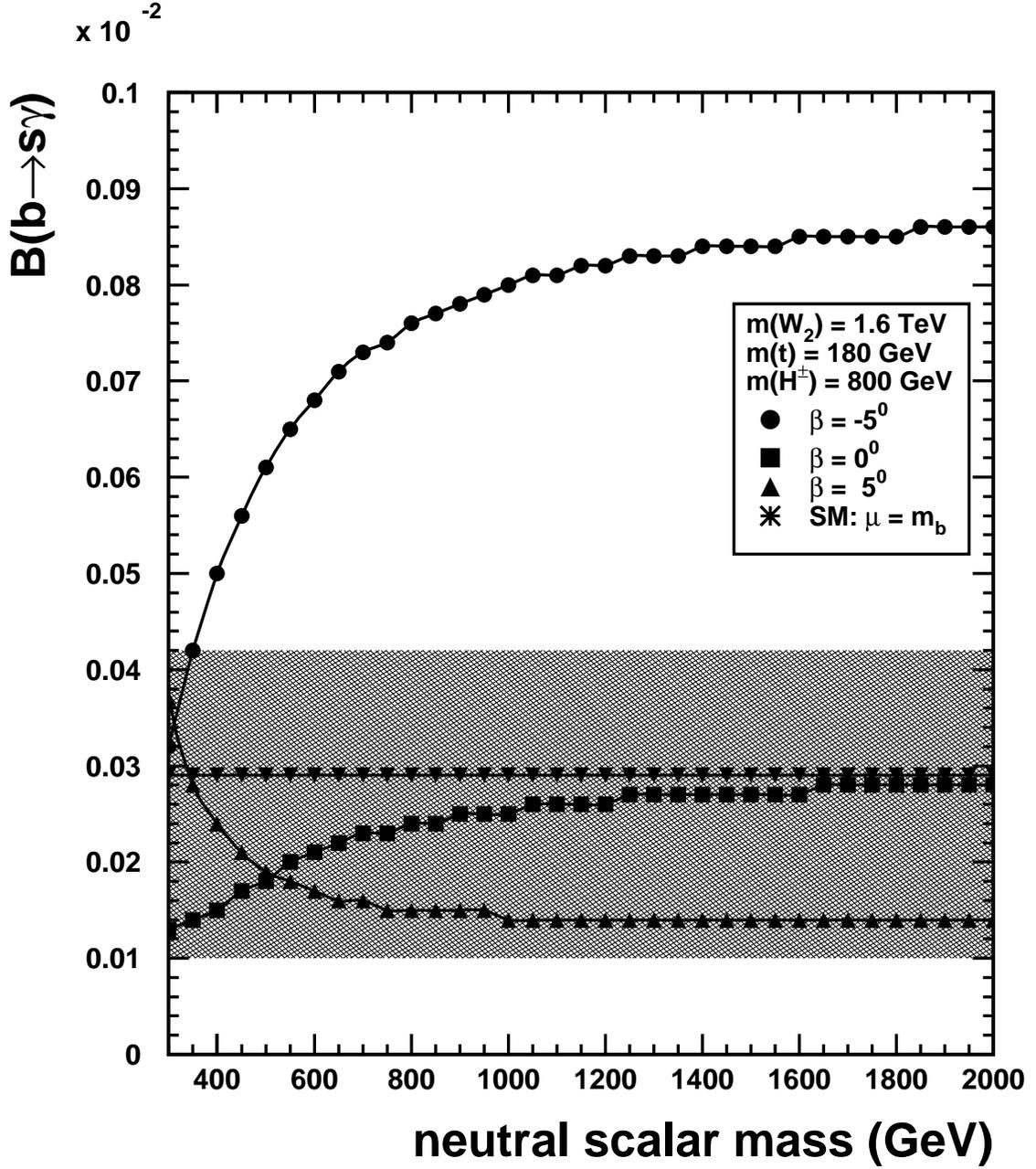}}
\vskip -2in\caption{The branching ratio for the process
$b\rightarrow s\gamma$ as a function of $m_{H_2^0}$
for different values of $\beta$ in the LRM. $H_2^0$ and
$G_2^0$ have been assumed to be mass-degenerate.
The value of $m_t$ has been fixed to 180 GeV
and $m_{H^\pm}$ has been set to 800 GeV. The SM line
relies on leading log QCD calculation at $\mu = m_b$. The shaded
area is the experimentally allowed region at 95\% C.L. [3].}
\end{figure}

\end{document}